\documentclass[twocolumn,twocolappendix]{aastex63}  

\hypersetup{linkcolor=red,citecolor=blue,filecolor=cyan,urlcolor=magenta}

\usepackage[normalem]{ulem}

\def\hi{\relax \ifmmode {\mbox H\,\textsc{i}}\else H\,{\scshape i}\fi}
\def\hii{\relax \ifmmode {\mbox H\,\textsc{ii}}\else H\,{\scshape ii}\fi}
\def\nii{\relax \ifmmode {\mbox N\,\textsc{ii}}\else N\,{\scshape ii}\fi}
\def\oi{\relax \ifmmode {\mbox O\,\textsc{i}}\else O\,{\scshape i}\fi}
\def\oii{\relax \ifmmode {\mbox O\,\textsc{ii}}\else O\,{\scshape ii}\fi}
\def\oiii{\relax \ifmmode {\mbox O\,\textsc{iii}}\else O\,{\scshape iii}\fi}
\def\cii{\relax \ifmmode {\mbox C\,\textsc{ii}}\else C\,{\scshape ii}\fi}
\def\ciii{\relax \ifmmode {\mbox C\,\textsc{iii}}\else C\,{\scshape iii}\fi}
\def\civ{\relax \ifmmode {\mbox C\,\textsc{iv}}\else C\,{\scshape iv}\fi}
\def\hei{\relax \ifmmode {\mbox He\,\textsc{i}}\else He\,{\scshape i}\fi}
\def\heii{\relax \ifmmode {\mbox He\,\textsc{ii}}\else He\,{\scshape ii}\fi}
\def\mgii{\relax \ifmmode {\mbox Mg\,\textsc{ii}}\else Mg\,{\scshape ii}\fi}
\def\sii{\relax \ifmmode {\mbox S\,\textsc{ii}}\else S\,{\scshape ii}\fi}
\def\siii{\relax \ifmmode {\mbox S\,\textsc{iii}}\else S\,{\scshape iii}\fi}
\def\hgi{\relax \ifmmode {\mbox Hg\,\textsc{i}}\else Hg\,{\scshape i}\fi}
\def\nai{\relax \ifmmode {\mbox Na\,\textsc{i}}\else Na\,{\scshape i}\fi}
\def\mgi{\relax \ifmmode {\mbox Mg\,\textsc{i}}\else Mg\,{\scshape i}\fi}
\def\caii{\relax \ifmmode {\mbox Ca\,\textsc{ii}}\else Ca\,{\scshape ii}\fi}

\received{2020 May 28}
\accepted{2020 September 20}

\submitjournal{ApJ}

\shorttitle{Local vs. global relations: $Z_g$ and $Age_\star$}
\shortauthors{S\'anchez-Menguiano et al.}

\begin{document}

\defcitealias{sanchezalmeida2019}{SA19}
\defcitealias{sanchezmenguiano2019}{SM19}

\title{Local and global gas metallicity versus stellar age relation in MaNGA galaxies}

\email{Laura.SanchezMenguiano@eso.org}

\author{Laura S\'anchez-Menguiano} 
\altaffiliation{ESO Fellow}
\affiliation{Instituto de Astrof\'isica de Canarias, La Laguna, Tenerife, E-38200, Spain}
\affiliation{Departamento de Astrof\'isica, Universidad de La Laguna, Spain}
\affiliation{European Southern Observatory, Karl-Schwarzschild-Str. 2, Garching bei M\"unchen, 85748, Germany}

\author{Jorge S\'anchez Almeida} 
\affiliation{Instituto de Astrof\'isica de Canarias, La Laguna, Tenerife, E-38200, Spain}
\affiliation{Departamento de Astrof\'isica, Universidad de La Laguna, Spain}

\author{Casiana Mu\~noz-Tu\~n\'on} 
\affiliation{Instituto de Astrof\'isica de Canarias, La Laguna, Tenerife, E-38200, Spain}
\affiliation{Departamento de Astrof\'isica, Universidad de La Laguna, Spain}

\author{Sebasti\'an F. S\'anchez}
\affiliation{Instituto de Astronom\'ia, Universidad Nacional Aut\'onoma de M\'exico, A.P. 70-264, C.P. 04510, M\'exico D.F., Mexico}

\begin{abstract}
The search for new global scaling relations linking physical properties of galaxies has a fundamental interest. Furthermore, their recovery from spatially resolved relations has been one of the spotlights of integral field spectroscopy (IFS). In this study we investigate the existence of global and local relations between stellar age ($Age_\star$) and gas-phase metallicity ($Z_g$). To this aim, we analyze IFS data for a sample of 736 star-forming disk galaxies from the MaNGA survey. We report a positive correlation between the global $Z_g$ and $D(4000)$ (an indicator of stellar age), with a slope that decreases with increasing galaxy mass. Locally, a similar trend is found when analyzing the $Z_g$ and $D(4000)$ of the star-forming regions, as well as the residuals resulting from removing the radial gradients of both parameters. The local laws have systematically smaller slopes than the global one. We ascribe this difference to random errors, that make the true slope of the $Age_\star-Z_g$ relation to be systematically underestimated when performing a least square fitting. The explored relation is intimately linked with the already known relation between gas metallicity and star formation rate at fixed mass, both presenting a common physical origin.
\end{abstract} 


\section{Introduction} \label{sec:intro}
The modeling of galaxies formed in a cosmological context has reached an impressive degree of realism \citep[e.g.,][]{ceverino2014, vogelsberger2014, vogelsberger2020, schaye2015, hopkins2014, hopkins2018, springel2018, nelson2019}. Model galaxies are able to reproduce many of the observed scaling relations (e.g., the distribution of masses, luminosities, and sizes, or the increase of star-formation rate and metallicity with mass), as well as properties of individual galaxies like the Milky Way or Andromeda \citep[e.g.,][]{nuza2014, scannapieco2015, carigi2019, grand2018, grand2019, buck2020}.  However, this successful modeling is not free from tune-up.  Simulations cannot resolve self-consistently all the relevant physical scales, from individual stars to cosmological volumes. Sub-grid physics is required to account for key processes like transforming gas into stars or super-nova feedback \citep[e.g.,][]{schaye2008, dallavecchia2012, marinacci2019, terrazas2020}. The free parameters encrypting such sub-grid physics are tuned to reproduce some of the observed scaling properties (e.g., the distribution of luminosities or stellar masses), whereas other scaling relations are used to evaluate the consistency of the simulated galaxies \citep[][]{crain2015, sanchezalmeida2018b, genel2018, torrey2019}. Testing against scaling relations represents the best benchmark available to judge the realism of the numerical simulations and, consequently, the way to assess our understanding on how galaxies form and evolve. Thus, investigations to disclose and characterize new galaxy scaling relations are essential. 

One scaling relation that has received notable attention is called {\em fundamental metallicity relation} \citep[FMR;][]{mannucci2010, laralopez2010b, ellison2008}. In general, more massive galaxies have larger gas-phase metallicity ($Z_g$) and larger star formation rate (SFR). However, this trend of increasing $Z_g$ with increasing SFR reverses when galaxies of the same stellar mass $M_\star$ are compared (provided $\log[M_\star/M_\odot] < 10.5$). The FMR shows that those galaxies with larger SFR have smaller $Z_g$. There is evidence for the FMR to hold until at least redshift 3 \citep[e.g.,][]{troncoso2014, sanchezalmeida2017}, but there is also dissenting views, mostly on whether the SFR is truly needed to describe the relation between $M_\star$ and $Z_g$ \citep[][]{izotov2014, delosreyes2015, sanders2015, barreraballesteros2017, sanchez2013, sanchez2017, sanchez2019b}. The FMR is taken as evidence for metal poor cosmic gas accretion fueling star formation \citep[e.g.,][]{mannucci2010, brisbin2012, dave2012}, an observationally-elusive but fundamental physical process according to numerical simulations \citep[e.g.,][]{dekel2009, sanchezalmeida2014}.

Often the global scaling relations, where each galaxy is a point, can be recovered from spatially resolved relations \citep[e.g.,][]{rosalesortega2012, sanchez2013, wuyts2013, barreraballesteros2016, canodiaz2016, hsieh2017, errozferrer2019}. \citet[][henceforth cited as SA19]{sanchezalmeida2019} show the FMR to result from the spatial integration of a local correlation between excess of $Z_g$ and excess of surface SFR. Indeed, \citet[][henceforth cited as SM19]{sanchezmenguiano2019} observe such local \mbox{(anti-)correlation} analyzing spatially resolved MaNGA galaxies, and similar conclusions are also drawn using other approaches and datasets \citep[][]{sanchezalmeida2018, hwang2019}. Given a number of simplifying  assumptions (e.g., small relative fluctuations),  \citetalias{sanchezalmeida2019} derive the mathematical equivalence between the global and the local correlation, which are predicted to have exactly the same logarithmic slope. Since it is a formal derivation, the relation still holds even when the variables are renamed, leading to the conclusion that the correspondence between local and global laws is not specific of the FMR.  This fact implies that there should be local counterparts associated with other known global scaling relations involving $Z_g$. In particular, the stellar age has been shown to be correlated with $Z_g$. For a set of local analogues to Ly-break galaxies, systems with younger stellar populations have lower $Z_g$ at fixed $M_\star$ \citep{lian2015}. The existence of this global correlation involving stellar age ($Age_\star$) suggests the existence of the corresponding local counterpart, namely, a local correlation between $Z_g$ and $Age_\star$.
The present Paper describes our work to disclose and characterize this foretold relation using the same MaNGA dataset employed in \citetalias{sanchezmenguiano2019}. 

The Paper is organized as follows: Section~\ref{sec:sample} briefly describes the MaNGA data and the galaxy sample, whereas the procedures to compute the physical parameters used in the analysis are given in Section~\ref{sec:analysis}. The derivation of the galaxy integrated correlation (global law) is included in Section~\ref{sec:results1}, with the local law worked out and characterized in Section~\ref{sec:results2}. Local and global correlations are compared in Section~\ref{sec:localglobal}. Finally, the results are discussed in Section~\ref{sec:conclusions}. Supplementary material includes Appendices~\ref{sec:appendix3},~\ref{sec:appendix1} and~\ref{sec:appendix2}. Appendix~\ref{sec:appendix3} presents several tests performed to assess the robustness of the results. Appendix~\ref{sec:appendix1} analyzes the $Z_g$ versus ${\rm Age}_\star$ relation when ages are computed from fitting the stellar continuum. Lastly, Appendix~\ref{sec:appendix2} shows how the slope of a relation is underestimated when using noisy data, a conclusion we employ to justify some of the results.

\section{Data and galaxy sample} \label{sec:sample}

\subsection{MaNGA data}\label{subsec:manga}
Mapping Nearby Galaxies at Apache Point Observatory \citep[MaNGA,][]{bundy2015} is an ongoing survey part of the fourth generation Sloan Digital Sky Survey (SDSS-IV). Its goal is to gather spatially resolved information of 10\,000 galaxies up to redshift $\sim0.15$ based on integral field spectroscopy (IFS) techniques. The data were collected using the BOSS spectrographs \citep{smee2013} mounted on the Sloan 2.5\,m telescope at Apache Point Observatory \citep{gunn2006}. The field of view (FoV) of the instrument varies from $12.5''$ to $32.5''$ in diameter for the five different hexagonal configurations displayed by the 17 simultaneous bundles of fibers \citep{drory2015}. The covered wavelength range spans from $3600$ \AA\ to $10300$ \AA, with a nominal resolution of $\lambda/\Delta\lambda \sim 2100$ at 6000 \AA\ \citep{smee2013}. 

The MaNGA mother sample consists of two main subsets of data: the {\it Primary} sample, comprising $\sim 5000$ galaxies observed up to $1.5$ effective radii ($R_e$); and the {\it Secondary} one, which includes $\sim 3300$ objects with a coverage up to $2.5\,R_e$ \citep{wake2017}. An additional third subsample, named the {\it Color-Enhanced} supplement and containing $\sim 1700$ galaxies, is selected to properly cover the underrepresented areas in the color-magnitude space (high-mass blue galaxies, low-mass red galaxies, and {\em green valley} galaxies).

The reduction of the MaNGA data is performed using an automatic pipeline \citep[the version used here is 2.4.3,][]{law2016} which includes standard steps such as bias subtraction and flat-fielding, flux and wavelength calibration, and sky subtraction. The resulting spectra for each sampled spaxel of $0.5'' \times 0.5''$ present a final spatial resolution of FWHM $\sim2.5''$, which corresponds to a physical resolution of $\sim 1.5$ kpc (at an average redshift of 0.03; see Section~\ref{subsec:sample} for details on the selected subsample).

Additional details on the MaNGA mother sample, survey design, observational strategy, and data reduction are provided in \citet{law2015}, \citet{yan2016}, \citet{law2016}, and \citet{wake2017}.

\subsection{Sample selection}\label{subsec:sample}
In this study we analyze a sample consisting of 736 star-forming galaxies with good physical resolution and spatial coverage extracted from the fifteenth MaNGA data released \citep{aguado2019}. Exactly the same sample was also used to study the local relation between SFR and $Z_g$ in \citetalias{sanchezmenguiano2019}. A complete description can be found in such article. Here we summarize the main criteria adopted to select the galaxies:
\begin{enumerate}
\item Galaxies with $z<0.05$ and observed with the largest FoVs ($27.5''$ and $32.5''$).
\item Galaxies with $b/a > 0.35$, that is, an inclination smaller than approximately 70$\degr$.
\item Galaxies with morphological types $T\geq1$, corresponding to types Sa and later, based on the classification carried out by \citet{fischer2019}.
\item Galaxies meeting the required quality standards of the data reduction pipeline (i.e., without bad flags in the DRP3QUAL field of the datacube FITS header, see \citealt{law2016} for details).
\item Galaxies containing at least 10 star-forming spaxels to properly characterize the local relation (see Section~\ref{sec:analysis} for details in the definition of these star-forming spaxels).
\end{enumerate}
\citetalias{sanchezmenguiano2019} proved that the galaxy properties resulting from this selection resemble those of the MaNGA mother sample with no obvious bias towards any particular subtype of star-forming galaxies. Additional tests discarding a major impact of the sample selection on the results are described in Appendix~\ref{sec:appendix3}.

\section{Analysis} \label{sec:analysis}
In order to derive the properties of the star-forming gas, we make use of the {\scshape Pipe3D} analysis pipeline \citep{sanchez2016a, sanchez2016b}, whose implementation for MaNGA data is described in detail in \citet{sanchez2019}. Briefly, first {\scshape Pipe3D} fits the stellar component using a linear combination of synthetic Single Stellar Population (SSP) templates, and subtracts it out from the original datacube to generate a pure gas cube. Then, the pipeline measures the emission line fluxes on the gas cube performing a multi-component fitting using both a single Gaussian function (per emission line and spectrum) and a weighted moment analysis. In addition to the flux intensity, {\scshape Pipe3D} obtains the equivalent width (EW), systemic velocity, and velocity dispersion for each of the 52 analyzed emission lines (including, for instance, H$\alpha$, H$\beta$, \mbox{[\oiii]~$\lambda5007$}, or \mbox{[\nii]~$\lambda6584$}).

The two-dimensional (2D) emission line intensity maps are then corrected for dust attenuation based on the extinction law from \citet{cardelli1989}, with $R_V=3.1$, and with the observed H$\alpha$/H$\beta$ Balmer decrement assumed to have a true value of 2.86 \citep{osterbrock1989}. To select star-forming regions (spaxels), we adopt the \citet{kewley2001} demarcation line on the \citet{baldwin1981} diagnostic diagram that involves the \mbox{[\nii]~$\lambda6584$/H$\alpha$} and \mbox{[\oiii]~$\lambda5007$/H$\beta$} line ratios. The spaxels located below such curve, and having an $\rm H\alpha$ equivalent width greater than $\rm 6 \,\AA$, are those associated with star formation. The latter criterion excludes low-ionisation sources \citep{cidfernandes2011}, and assumes that a significant percentage of the emission of the star-forming regions is produced by young stars \citep[which induces high H$\alpha$ equivalent width, see e.g.][]{sanchez2014}. Finally, spaxels with a signal-to-noise ratio (SNR) lower than 3 in any of the emission lines involved in the derivation of the oxygen abundances (see Section~\ref{subsec:properties}) are discarded from further analysis. The SNR is estimated from the relative error of the line flux intensities (i.e., the ratio of the flux to the flux error). We note that using a SNR threshold of 1 instead of 3 does not seem to significantly affect the results (see Appendix~\ref{sec:appendix3} for details).

\begin{figure*}
	\begin{center}
		\resizebox{\hsize}{!}{\includegraphics{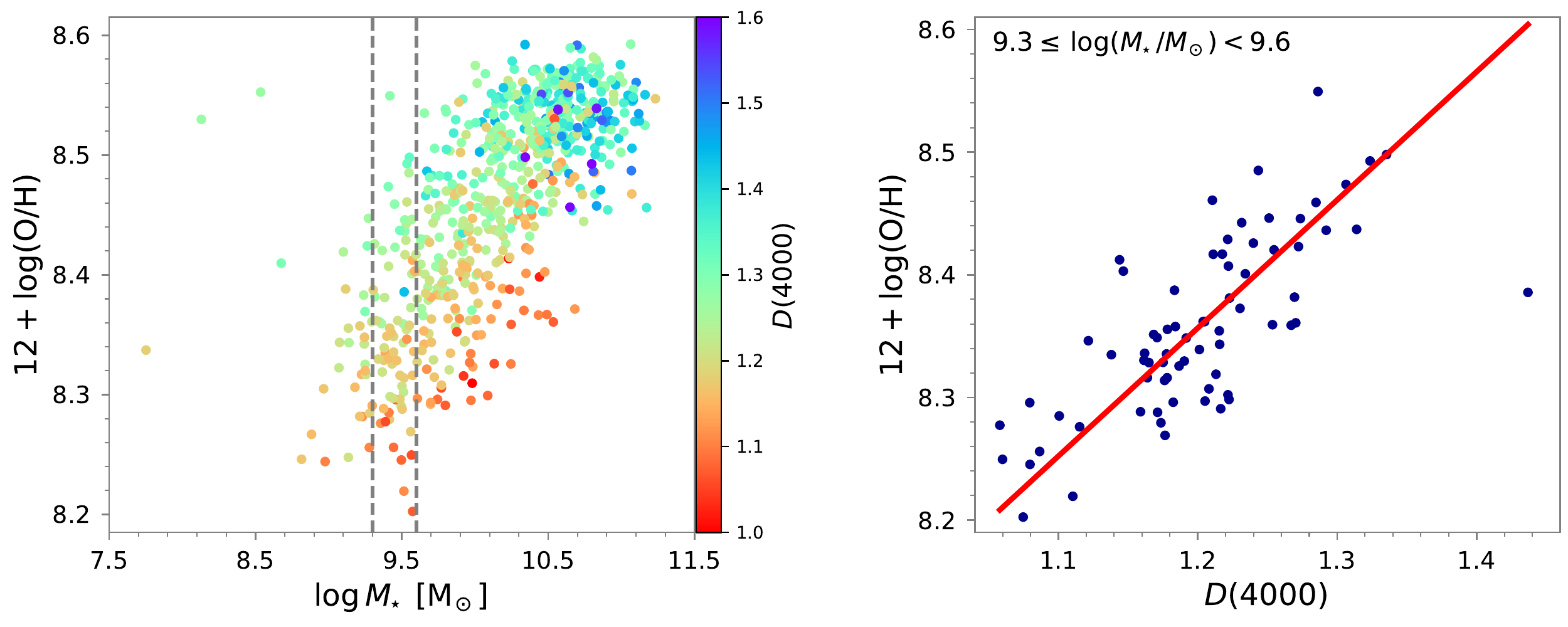}}
		\caption{{\it Left panel}: Global MZ relation color-coded with $D(4000)$. Given $M_\star$, galaxies with larger metallicity tend to have larger $D(4000)$ (i.e., older stellar ages). {\it Right panel}: scatter plot gas metallicity versus strength of the 4000-\AA \,break for the galaxies in the mass bin $9.3 \leq \log (M_\star/M_\odot) < 9.6$, which is marked with dashed lines in the left panel. The red solid line represents the ODR fitting of the data points.}
		\label{fig1}
	\end{center}
\end{figure*}

\subsection{Derivation of galaxy properties}\label{subsec:properties}
In this section, we describe the procedures to derive all the parameters analyzed along this study, namely, the gas metallicity, the stellar mass, and the line-strength index $D(4000)$. As a proxy for $Z_g$, we measure the oxygen abundance (O/H) of the selected star-forming spaxels adopting the empirical calibration for the O3N2 index proposed by \citet{marino2013}:
\begin{equation}
12+\log\left({\rm O/H}\right) = 8.533 - 0.214 \,\times\, {\rm O3N2}
,\end{equation}
with $\rm O3N2 =  \log\left([\oiii] \lambda5007/H\beta \times H\alpha/[\nii] \lambda6584\right)$. The calibration error associated with the scatter in the relation is 0.08 dex \citep[see][]{marino2013}. It constitutes one of the most accurate calibrations to date for the O3N2 index, especially in the high-metallicity regime, where previous calibrators lack high quality observations \citep[e.g.][]{pettini2004, perezmontero2009}. Furthermore, this abundance indicator was successfully employed by \citetalias{sanchezmenguiano2019}, where it was shown to be fully consistent with other methods based on photo-ionization models \citep{perezmontero2014}.

In addition, $M_\star$ is estimated by co-adding the stellar surface mass density ($\Sigma_\star$) of the spaxels. The $\Sigma_\star$ values are measured by {\scshape Pipe3D} from the SSP model spectra adopting a Salpeter IMF \citep{sanchez2016b}. 

Finally, $D(4000)$, that is, the index parametrizing the strength of the 4000-\AA \, break, is derived by the algorithm as the ratio between the integrated flux in the narrow bandpasses $4050-4250$ \AA \, and $3750-3950$ \AA. These fluxes are measured in the spectra once the strong emission lines have been subtracted \citep[for more details, see sec.~3.6.1. of][]{sanchez2016b}. When quoting galaxy integrated values, we use the average $D(4000)$ of all analysed star-forming regions of the entire galaxy. We note that deriving the global $Z_g$ and $D(4000)$ from the characteristic values at $R_e$ does not affect the conclusions presented in this work (see Appendix~\ref{sec:appendix3} for more information).

\section{The global $Age_\star-Z_g$ relation}\label{sec:results1}

A global relation between $D(4000)$ \citep[as an indicator for the galaxy stellar age, e.g.][]{kauffmann2003, gallazzi2005, sanchezalmeida2012} and $Z_g$ was found by \citet{lian2015}, according to which the galaxies of the same $M_\star$ that are more metal rich also present older stellar ages. As a proxy for the gas metallicity, the authors used two different empirical calibrations of oxygen abundances: the one by \citet{pettini2004} based on the N2 index (defined as $\rm N2 = \log ([\nii]~\lambda6584/H\alpha)$), and the one proposed in \citet{mannucci2010} based on the N2 to R23 ratio (the latter defined as $\rm R23 = \log (([\oii]~\lambda3727+[\oiii]~\lambda\lambda4959, 5007)/H\beta)$).

In order to assess the existence of this relation in our galaxy sample, one would ideally compare precise stellar age estimations with those of gas metallicity. However, the derivation of stellar ages from integrated spectroscopy is far from straightforward, involving complicated inversion methods and relying (and thus depending) on stellar modelling. In an attempt to simplify the characterization of stellar ages, here we focus on direct measurements of the $D(4000)$ line-strength index, as \citet{lian2015} did. Nevertheless, we note that similar trends (in both, global and local approaches) are obtained from spectral fitting using Pipe3D (details are given in Appendix~\ref{sec:appendix1}). Regarding $Z_g$, we use an indicator based on the O3N2 method (Section~\ref{subsec:properties}). 

\begin{figure*}
\begin{center}
\resizebox{\hsize}{!}{\includegraphics{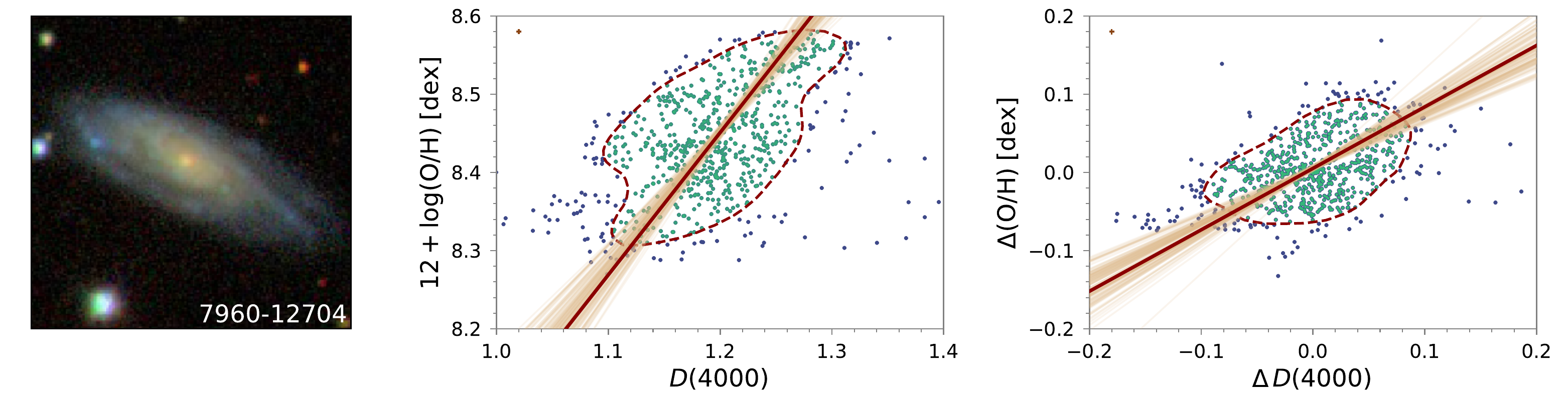}}
\caption{Scatter plots representing the original (middle panel) and residual (right panel) local relations between the gas metallicity and the \mbox{4000-\AA} break for one typical galaxy of the sample, 7960-12704. Its SDSS color image is shown in the left panel. The red solid lines correspond to an ODR fitting to the data points enclosed within the $80\%$ density-contour indicated by the red dashed lines, and the shaded areas show the error region estimated by bootstrapping (see main text). The error bars in the top left corner of the panels correspond to the average error of all analyzed spaxels. Note that the calibration error associated with the use of the O3N2 metallicity indicator is not included.} 
\label{fig2}
\end{center}
\end{figure*}

The left panel of Figure~\ref{fig1} represents the global mass-metallicity relation (hereafter MZ relation) color-coded according to $D(4000)$. Every galaxy is a point. There is a global trend for more metallic (more massive) objects to have larger $D(4000)$ values (i.e., to be older). On top of this, we clearly see that at a fixed mass (for instance, the mass bin $9.3 \leq \log (M_\star/M_\odot) < 9.6$, marked with grey dashed lines), galaxies of larger $Z_g$ are characterized by larger $D(4000)$. This is especially evident for less massive galaxies ($\log (M_\star/M_\odot) \lesssim 10.5$). In the right panel of Figure~\ref{fig1} we show the scatter plot $Z_g$ (in logarithmic scale) versus $D(4000)$ for the galaxies in the above mentioned mass bin ($9.3 \leq \log (M_\star/M_\odot) < 9.6$). A clear positive correlation between the gas metallicity and the stellar age is observed. An orthogonal distance regression (ODR) fitting has been performed to the data points and the result is represented by the red solid line. Contrary to ordinary linear regression, where the goal is to minimize the sum of the squared vertical distances between the measured and fitted $y$ values, ODR minimizes the perpendicular distances from the data points to the regression line \citep{adcock1878}. This technique is recommended when the errors of both analysed variables are comparable, and therefore frequently used in the literature, for instance, to relate galaxy properties \citep[e.g.][]{hsieh2017, bisigello2018, colombo2019, lin2019, barreraballesteros2020}. Performing the ODR fitting to $Z_g$ versus $D(4000)$ for galaxies in seven different mass bins of 0.3-width from 9.0 to 11.1 yield positive correlations with a slope that decreases when increasing galaxy mass, until the sign reverses for masses larger than $\rm \log (M_\star/M_\odot) \sim 11$. In all cases except the last two, the correlation coefficient is larger than 0.6, with a p-value below $1\%$, indicating a clear correlation between both parameters that tends to disappear when moving to high masses. The whole trend will be shown and discussed in Section~\ref{sec:localglobal}. The gas metallicity, stellar mass, and $D(4000)$ of each galaxy in the sample are compiled in Table~\ref{table1}. 

\section{The local $Age_\star-Z_g$ correlation}\label{sec:results2}

To our knowledge, no previous investigation on the local \mbox{$Age_\star-Z_g$} relation has been carried out so far. For the first time, we seek for such correlation and its connection with the galaxy-integrated trend linking both parameters. 

For this local approach, one has to bear in mind that the 2D maps of all stellar properties derived with {\scshape Pipe3D}, including $D(4000)$, present a spatial binning beyond the spaxel size, which was adopted to increase the SNR of the spectra so as to obtain an accurate estimation of the stellar contribution. The typical size of the bins range between 2 and 5 spaxels in most cases, with a few larger ones in the outer regions of the galaxies \citep[for more information on this spatial binning and its implementation in MaNGA, see][]{sanchez2019}. However, gas properties such as $Z_g$ are estimated in every single spaxel. In order to reconcile the two different scales, all the star-forming spaxels belonging to the same bin are considered as a single data point, with its $Z_g$ defined as the average value within the bin and its $D(4000)$ as the value derived by {\scshape Pipe3D} from the fitting of the co-added spectra within the bin. 

\begin{figure*}
\begin{center}
\resizebox{\hsize}{!}{\includegraphics{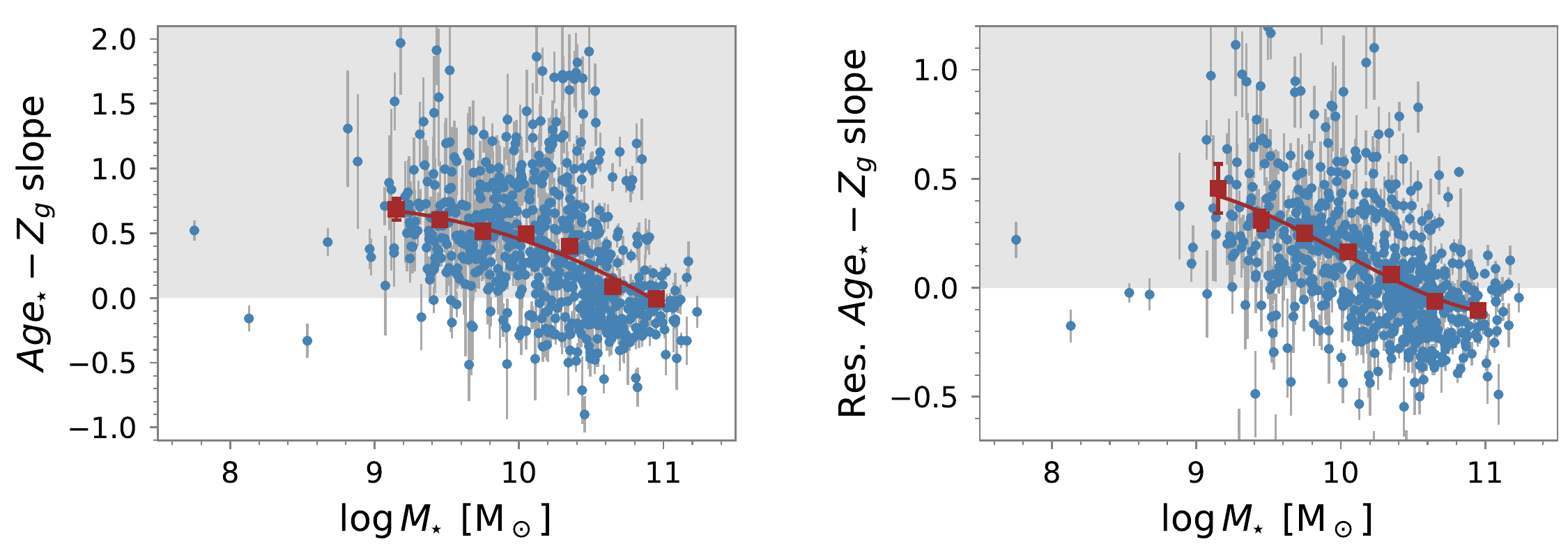}}
\caption{Variation with galaxy mass of the slope of the original (left panel) and residual (i.e., after removing the radial gradients of both parameters; right panel) $Age_\star-Z_g$ local relations. The dark-red squares represent the mean slope in 7 stellar mass bins, with the error bars indicating the error given by $\sigma_{\rm bin}/\sqrt{N_{\rm bin}}$ ($\sigma_{\rm bin}$ is the standard deviation of the slopes in each bin and $N_{\rm bin}$ the number of galaxies within). In most cases, the error bars are smaller than the marker size. The average values (dark-red squares) were fitted with a third-order polynomial (dark-red solid lines).} 
\label{fig3}
\end{center}
\end{figure*}

In order to describe the local relation between these quantities, we follow two different approaches: (1) to include their radial decline when moving from the inner to the outermost parts of galaxy disks, and (2) to describe their local variations once the overall radial profiles have been removed. Hereafter, we will refer to them as original and residual local relations, respectively. For the original relation, we examine for each galaxy the scatter plot $D(4000)$ versus $Z_g$ for all the star-forming bins. For the residual relation, we analyze the correlation between the residuals. These residuals are obtained by subtracting the azimuthally averaged radial $Z_g$ and $D(4000)$ distributions from the 2D maps. These averaged values are measured in elliptical annuli centred at each data point, taking into account the position angle and ellipticity of the galaxies to correct for inclination. The width of the annuli is $2''$, which is comparable to the spatial resolution of the data (see Section~\ref{subsec:manga}).

Figure~\ref{fig2} shows scatter plots representing the original (left) and residual (right) $Age_\star-Z_g$ local relations derived for one typical galaxy of the sample, 7960-12704. The average error of the analyzed star-forming regions is displayed on the top left corner of each panel. We note that the calibration error of the O3N2 metallicity indicator (0.08 dex) is not included. For the particular galaxy shown in Figure~\ref{fig2}, positive correlations can be observed between both the original (left panel) and residual (right panel) variations of $Z_g$ and $D(4000)$. In order to quantify the correlations, we perform an ODR fitting to the data points enclosed within the $80\%$-level density contour (i.e., the contour that encircles $80\%$ of the total number of data points; red dashed line). The error in the linear fit (brown shaded area) is estimated by bootstrapping. We show the standard deviation of the slopes in 100 fits inferred from a random re-sampling of the data with replacement (i.e. each data point can appear multiple times in the sample). The root-mean-square error (RMSE) of the fit is 0.033 for the original relation (left), and 0.028 for the residual relation (right). On average, galaxies in the sample present a slightly smaller RMSE, with median values of 0.024 and 0.018, respectively. This scatter is similar to the one found for the local relation between SFR and $Z_g$ described in \citetalias{sanchezmenguiano2019}.

\begin{deluxetable*}{l@{\hspace{1cm}}CCCCC}
\tablecaption{Global and local parameters of the individual galaxies. \label{table1}}
\tablenum{1}
\tablehead{
\colhead{Galaxy ID} & \colhead{$\log M_\star$} & \colhead{$Z_g$} & \colhead{$D(4000)$} & \colhead{$Age_\star-Z_g$ slope} & \colhead{$Age_\star-Z_g$ slope} \\
\colhead{} & \colhead{[M$_{\odot}$]} & \colhead{[dex]} & \colhead{} & \colhead{(original)} & \colhead{(residual)}
}
\colnumbers
\startdata
7443-12703  &  11.07  &  8.47  &  1.16  & $ -0.07 \pm 0.04 $ & $ +0.09 \pm 0.03 $ \\ 
7443-9101  &  10.68  &  8.48  &  1.16  & $ +0.47 \pm 0.09 $ & $ +0.52 \pm 0.09 $ \\ 
7495-12704  &  10.72  &  8.58  &  1.33  & $ -0.23 \pm 0.07 $ & $ -0.24 \pm 0.03 $ \\ 
7495-9101  &  9.18  &  8.31  &  1.16  & $ +2.0 \pm 0.4 $ & $ +1.7 \pm 0.3 $ \\ 
7815-12701  &  9.43  &  8.33  &  1.19  & $ +0.38 \pm 0.15 $ & $ +0.34 \pm 0.12 $ \\ 
7815-12702  &  9.65  &  8.37  &  1.26  & $ +1.1 \pm 0.3 $ & $ -4.9 \pm 0.3 $ \\ 
7815-12704  &  10.87  &  8.54  &  1.42  & $ -0.14 \pm 0.07 $ & $ -0.23 \pm 0.06 $ \\ 
7815-9101  &  10.21  &  8.51  &  1.27  & $ +1.2 \pm 0.2 $ & $ -0.13 \pm 0.09 $ \\ 
7815-9102  &  9.66  &  8.38  &  1.19  & $ +0.4 \pm 0.6 $ & $ +0.17 \pm 0.13 $ \\ 
7957-12702  &  10.50  &  8.45  &  1.23  & $ +1.03 \pm 0.09 $ & $ +0.17 \pm 0.04 $ \\ 
7957-12704  &  10.16  &  8.45  &  1.25  & $ +0.60 \pm 0.11 $ & $ +0.19 \pm 0.05 $ \\ 
7957-9101  &  9.77  &  8.38  &  1.18  & $ +0.9 \pm 0.2 $ & $ +0.13 \pm 0.13 $ \\ 
7957-9102  &  10.30  &  8.51  &  1.20  & $ +1.70 \pm 0.17 $ & $ +0.19 \pm 0.04 $ \\ 
7958-12701  &  9.61  &  8.30  &  1.12  & $ +0.3 \pm 0.2 $ & $ +0.11 \pm 0.08 $ \\ 
7958-12703  &  10.18  &  8.48  &  1.25  & $ +0.30 \pm 0.14 $ & $ -0.01 \pm 0.11 $ \\ 
7958-12705  &  9.82  &  8.32  &  1.21  & $ +0.47 \pm 0.12 $ & $ +0.28 \pm 0.07 $ \\ 
7960-12701  &  11.01  &  8.56  &  1.26  & $ -0.24 \pm 0.07 $ & $ -0.35 \pm 0.04 $ \\ 
7960-12703  &  10.24  &  8.52  &  1.29  & $ +1.7 \pm 0.2 $ & $ +0.14 \pm 0.07 $ \\ 
7960-12704  &  10.40  &  8.44  &  1.20  & $ +1.82 \pm 0.14 $ & $ +0.79 \pm 0.07 $ \\ 
7960-12705  &  10.51  &  8.57  &  1.33  & $ -0.19 \pm 0.17 $ & $ +0.02 \pm 0.04 $ \\ 
7962-12701  &  9.96  &  8.46  &  1.17  & $ +1.1 \pm 0.2 $ & $ +0.17 \pm 0.08 $ \\ 
7962-12702  &  10.38  &  8.45  &  1.21  & $ +1.7 \pm 0.2 $ & $ +0.38 \pm 0.07 $ \\ 
7962-12703  &  11.12  &  8.53  &  1.42  & $ -0.33 \pm 0.06 $ & $ -0.11 \pm 0.04 $ \\ 
7962-12704  &  10.16  &  8.52  &  1.28  & $ -0.38 \pm 0.12 $ & $ -0.05 \pm 0.03 $ \\ 
7964-12701  &  9.51  &  8.36  &  1.27  & $ +0.58 \pm 0.18 $ & $ +0.60 \pm 0.12 $ \\ 
7968-12702  &  10.44  &  8.52  &  1.30  & $ +0.38 \pm 0.09 $ & $ +0.17 \pm 0.05 $ \\ 
7972-12703  &  10.36  &  8.53  &  1.34  & $ -0.28 \pm 0.08 $ & $ -0.11 \pm 0.06 $ \\ 
7972-12704  &  10.10  &  8.51  &  1.31  & $ +1.2 \pm 0.2 $ & $ +0.08 \pm 0.08 $ \\ 
7975-12705  &  9.77  &  8.31  &  1.17  & $ +1.05 \pm 0.18 $ & $ +0.20 \pm 0.14 $ \\ 
7975-9102  &  10.31  &  8.49  &  1.27  & $ +0.57 \pm 0.12 $ & $ +0.11 \pm 0.06 $ \\ 
7977-9101  &  11.23  &  8.55  &  1.18  & $ -0.11 \pm 0.12 $ & $ -0.05 \pm 0.07 $ \\ 
7990-12701  &  10.19  &  8.54  &  1.29  & $ +0.20 \pm 0.17 $ & $ -0.40 \pm 0.10 $ \\ 
7990-12704  &  10.51  &  8.53  &  1.39  & $ -0.13 \pm 0.09 $ & $ -0.22 \pm 0.07 $ \\ 
7990-9101  &  10.20  &  8.55  &  1.38  & $ -0.3 \pm 0.2 $ & $ -0.44 \pm 0.15 $ \\ 
7991-12701  &  10.65  &  8.48  &  1.16  & $ +0.93 \pm 0.11 $ & $ +0.29 \pm 0.04 $ \\ 
7991-12704  &  9.83  &  8.50  &  1.25  & $ +0.8 \pm 0.2 $ & $ +0.23 \pm 0.06 $ \\ 
7991-9101  &  10.14  &  8.54  &  1.31  & $ +0.09 \pm 0.10 $ & $ -0.25 \pm 0.06 $ \\ 
7992-9101  &  9.44  &  8.30  &  1.21  & $ +0.6 \pm 0.3 $ & $ +1.89 \pm 0.19 $ \\ 
\enddata
\tablecomments{From left to right the columns correspond to (1) galaxy ID, defined as the [plate]-[ifudesign] of the MaNGA observations; (2) galaxy stellar mass; (3) gas-phase oxygen abundance; (4) 4000\,\AA-break index; (5) slope of the original local $Age_\star-Z_g$ relation; and (6) slope of the residual local $Age_\star-Z_g$ relation.\\[0.1cm] (This table is available in its entirety in machine-readable form only.)}
\end{deluxetable*}

The local relations between $D(4000)$ and $Z_g$ are derived for all the galaxies in the sample, and the resulting slopes of the linear fits, together with their corresponding errors, are listed in Table~\ref{table1}. In order to have sufficient statistics for the fits, 68 galaxies containing less than 30 data points (i.e., bins) are discarded from the analysis. We find that 65\% of the remaining 668 galaxies display positive correlation (i.e., positive slope) within the error bars, 20\% of them present anti-correlation (i.e., negative slope), and the remaining 15\% are compatible with absence of correlation (i.e., zero slope). When considering the residuals $\Delta D(4000)$ and $\Delta Z_g$, 47\% (31\%) of the galaxies show positive (negative) correlation, with 22\% of the sample showing lack of correlation within the observational errors. 

To investigate how the local relation between $D(4000)$ and $Z_g$ depends on the galaxy mass, Figure~\ref{fig3} shows the slope of each galaxy as a function of the stellar mass for the original (left) and residual (right) relations. Overall, we can see that lower mass galaxies present a local correlation between the gas metallicity and the stellar age with a larger slope than more massive galaxies, where the correlation tends to disappear. This behavior is shown by the red solid lines, which represent the variation of the mean slope with $M_\star$. In the case of the residual local relation (right panel in Figure~\ref{fig3}), the derived slopes are a bit smaller, but differ significantly from zero. There seems to exist the same tendency shown by the original local relation (left panel in Figure~\ref{fig3}) of the least massive galaxies presenting the largest slopes. Moreover, in this case the relation reverses at  $\rm \log (M_\star/M_\odot) \sim 10.5$, with more massive galaxies presenting an anti-correlation between the residuals $\Delta D(4000)$ and $\Delta Z_g$, that is, the older the less metallic.

\section{Linking the global and local trends}\label{sec:localglobal}
As explained in Section~\ref{sec:intro}, \citetalias{sanchezalmeida2019} derive an analytical formulation to draw global laws from local ones. As an example, they show how the global FMR emerges from the local correlation between SFR surface density and $Z_g$ found by \citetalias{sanchezmenguiano2019}. In this section we study the correspondence between the local and global trends for the $Z_g$ versus $D(4000)$ relation described in Sections~\ref{sec:results1} and \ref{sec:results2}. 

\begin{figure}
\begin{center}
\resizebox{\hsize}{!}{\includegraphics{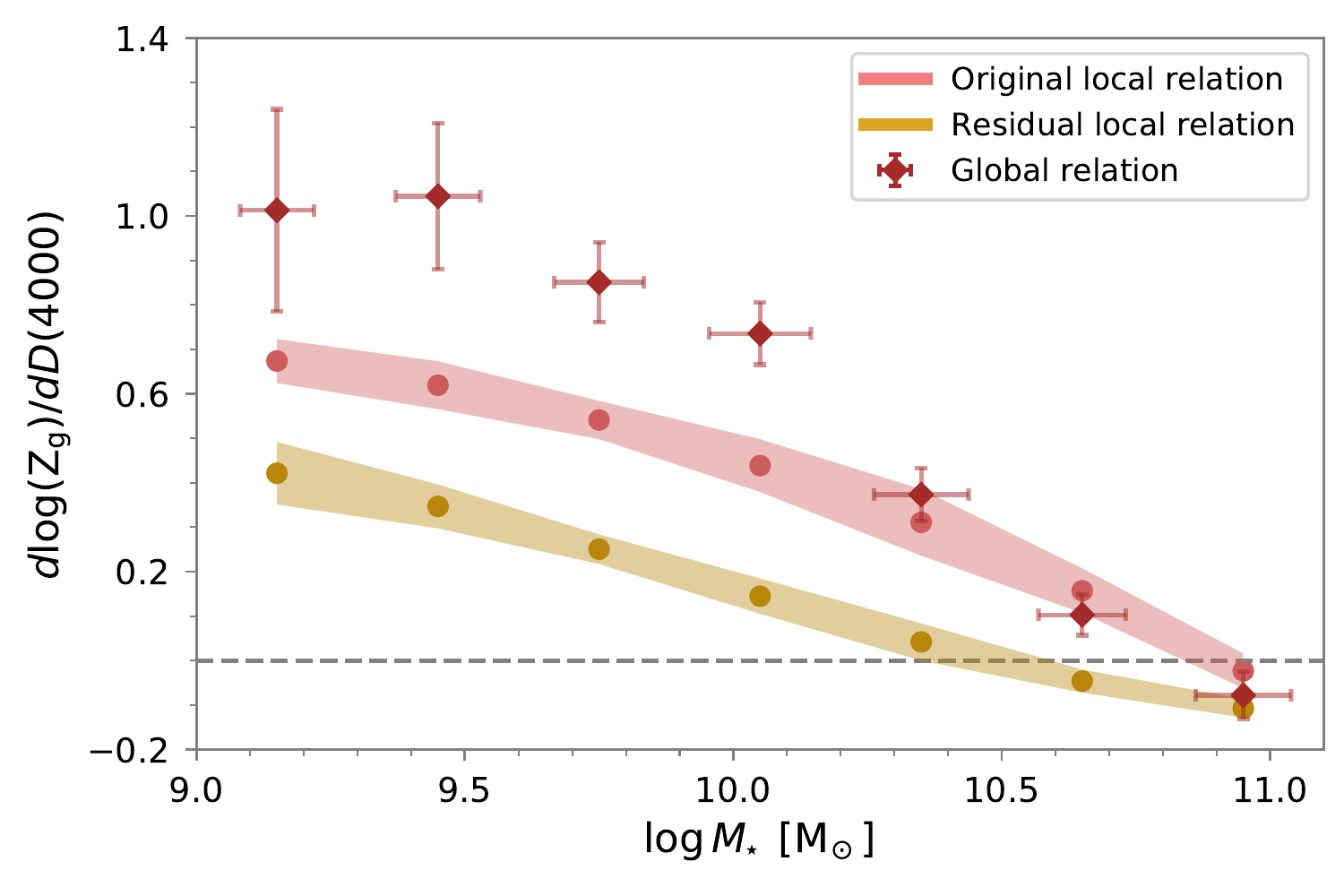}}
\caption{Comparison of the slopes inferred from the local and global $Age_\star \!-Z_g$ relations as function of $M_\star$. The pink and yellow markers represent the mean slopes for the different mass bins obtained from spatially resolved data for the original and residual local relations, respectively. The error of these quantities are denoted by the shaded areas. The diamonds show the slopes derived from galaxy-integrated values using exactly the same MaNGA sample, with the error bars indicating the standard deviation in 100 fits inferred by bootstrapping (see main text).} 
\label{fig4}
\end{center}
\end{figure}

Figure~\ref{fig4} shows the comparison of the slopes inferred from the global and local relations between Z$_g$ and $D(4000)$ at each M$_\star$. For the global trend (diamonds), the error bars represent the standard deviation of the slopes derived in each mass bin from the integrated Z$_g$ and $D(4000)$ values in 100 fits inferred by bootstrapping. For the local trend, we represent both the original relation (pink markers) and the residual one (yellow markers). In this case, the shaded areas denote the error of the mean slope of the individual galaxies within each mass bin, as indicated by the error bars in Figure~\ref{fig3}. We can see that the three relations show the same trend, that is, a decreasing slope with increasing galaxy mass. In spite of the similarity in the tendency, the relations do not agree quantitatively; the global relation has slopes larger than the local ones in the low mass regime, with the residual local relation presenting the values closest to zero. The explanation for the qualitative agreement and the seeming quantitative discrepancy is discussed in Section~\ref{sec:conclusions}.

\section{Discussion and conclusions}\label{sec:conclusions}
The search for relations between integrated physical properties in galaxies has been widely addressed in the literature. With the advent of IFS, the availability of spectra providing spatially resolved information for large samples of galaxies has shown that many of these global trends emerge from local correlations. An example is the FMR, which is claimed to arise from the local anti-correlation between surface density SFR and $Z_g$ found by \citetalias{sanchezmenguiano2019} \citep[see also][for previous findings on dwarf galaxies]{sanchezalmeida2018}. The mathematical correspondence between both relations described by \citetalias{sanchezalmeida2019} predicts that the same global-local equivalence should hold for other parameters.

In this study we analyze IFS data for a sample of 736 star-forming disk galaxies from the MaNGA survey. The aims of this analysis are (i) to explore the existence of a local counterpart for a specific global relation, (ii) to investigate how the global and local trends compare with each other, and (iii) to assess the validity of the prediction given in \citetalias{sanchezalmeida2019} for the mathematical equivalence between the analyzed global law and the corresponding local correlation.

The examined relation links the gas metallicity and the stellar age. At a fixed stellar mass, galaxies with younger stellar populations (characterized by larger $D(4000)$ values) present lower $Z_g$ \citep{lian2015}. This trend is explained as a natural result of passive evolution: galaxies with older stellar populations experience a stronger metal enrichment process and thus exhibit larger gas metallicities. It may also be due to recent metal-poor gas accretion, which simultaneously decreases $Z_g$ and triggers star-formation, thus decreasing the mean light-weighted age of the stellar population. Which of the two processes dominate is expected to be a function of $M_\star$. Dividing the sample in seven mass bins (from 9.0 to 11.1 $\rm \log M_\odot$), we systematically find a clear positive correlation between global $Z_g$ (expressed as the oxygen abundance measured with the O3N2 indicator) and $D(4000)$, in agreement with \citet{lian2015}. The slope of the correlation decreases when increasing the mass of the bin and becomes zero at about $\rm \log (M_\star/M_\odot) \sim 11$.  

We characterize the local relation between $Z_g$ and $D(4000)$ within each galaxy by the slope of an ODR fitting to the scatter plot $\log Z_g$ versus $D(4000)$ after filtering out extreme values (this was called the {\it original local relation}). For the 668 galaxies with reliable fits, we find that 65\% present positive correlation, 20\% negative correlation, and 15\% are compatible within errors with no correlation between $Z_g$ and $D(4000)$. The positive correlation observed for most galaxies is expected as a consequence of the negative radial gradients found for both the gas metallicity \citep[e.g.][]{sanchez2014, ho2015, sanchezmenguiano2016, belfiore2017} and the stellar age \citep[e.g.][]{sanchezblazquez2014, gonzalezdelgado2015, goddard2017, ruizlara2017}, which are explained in the framework of an inside-out formation of the galaxy disks \citep{prantzos2000}. The presence of deviations from these radial declines for both parameters \citep[e.g.][]{ruizlara2016, sanchezmenguiano2018} might correspond to the cases where negative or lack of correlation between $Z_g$ and $D(4000)$ has been reported. However, this cannot be the full explanation since the correlations remain when the radial variations are removed. Moreover, there is a systematic change of the slope with galaxy stellar mass, positive at low mass and negative in the high mass end.    

We also study the local relation between the residual $\Delta Z_g$ and $\Delta D(4000)$ for all the star-forming regions obtained after removing the radial gradients (this is called the {\it residual local relation}). Deriving the slope of this correlation for the galaxy sample, we find that 47\% exhibit positive correlation, 31\% negative correlation, and the remaining 22\% lack of correlation between $\Delta Z_g$ and $\Delta D(4000)$. These results confirm that similarly to the global trend, in general older star-forming regions exhibit larger gas metallicities compared to younger regions within the same galaxy. 

For both the original and the residual $Age_\star-Z_g$ local relations, we find the slope to depend on $M_\star$, so that less massive galaxies are characterized by larger slopes than more massive systems. In the first case, the correlation disappears at about $\rm \log (M_\star/M_\odot) \sim 11$, whereas this happens at $\rm \log (M_\star/M_\odot) \sim 10.5$ for the residual local relation, reversing the relation for galaxies more massive than that. Being the stellar age a proxy for the shape of the star formation history (SFH) of a galaxy, this result indicates that the SFH of older and more massive regions present little variation among the average. However, at low mass we find a wide range of SFHs \citep[and chemical enrichment; e.g.][]{ibarramedel2019}.

The trend of the slope of the local relations with galaxy mass resembles that of the global law. As we discuss in Section~\ref{sec:intro}, every global correlation should come together with a local counterpart with the slopes of the global and local laws expected to be the same. However, we find the local correlations to have systematically smaller slopes than the global correlation. This is clear in the relation between excess of metallicity and excess of age (Figure~\ref{fig4}). The largest discrepancies are found for low-mass galaxies, where the slopes of the residual local relation are as low as half the slope of the global relation (although the error bars associated with the global slopes are also larger in this mass range, which could account for part of the discrepancies). A much better agreement is found for high-mass galaxies, where the differences are reduced by more than half. These differences do not disappear when deriving stellar ages from SSP fitting (see Appendix~\ref{sec:appendix1}). Indeed, although similar trends are observed for the mass-dependence of both the global and local relations, the reported slopes of the relations are significantly lower than when using $D(4000)$. This is especially noticeable for the residual local relation, where the percentage of slopes compatible with zero has significantly increased (from 22\% to 45\%), making the differences between the global and local relations more pronounced. However, we think that the uncertainties involved in the inversion method and the complex modelling of the stellar populations applied to spatially resolved IFS data \citep[e.g.][]{cidfernandes2014, sanchez2016a, bittner2019} might be partially responsible for these larger discrepancies. As we argue below, the clearer $Age_\star-Z_g$ relations reported when using $D(4000)$ do not result from the possible influence of other physical parameters on $D(4000)$ but describes a true relation between gas metallicity and stellar age.

\begin{figure}
	\begin{center}
		\resizebox{\hsize}{!}{\includegraphics{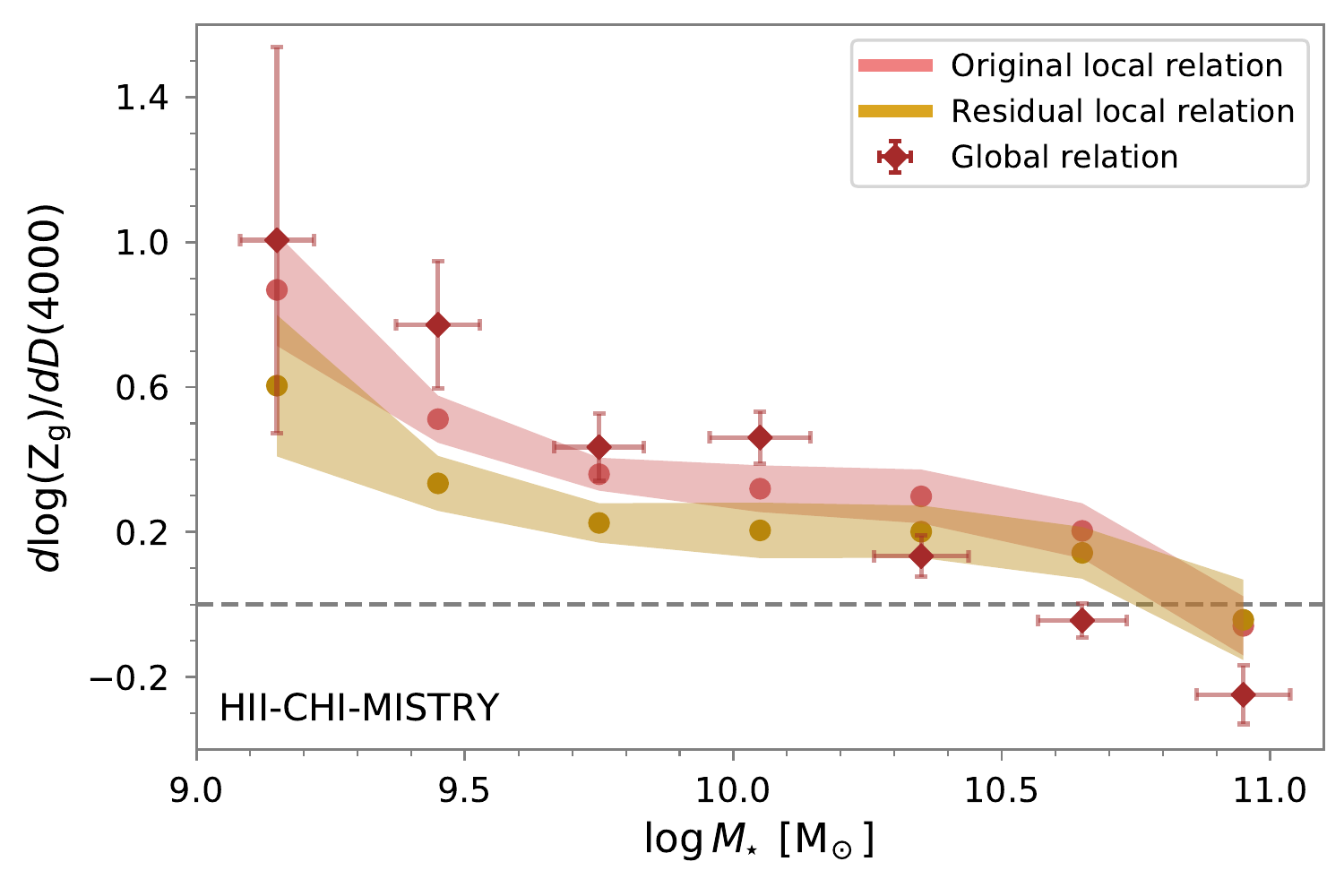}}
		\caption{Comparison of the slopes inferred from the local and global $Age_\star \!-Z_g$ relations as function of $M_\star$. $Z_g$ is determined using the HII-CHI-MISTRY calibration. See caption of Figure~\ref{fig4} for more details.} 
		\label{fig5}
	\end{center}
\end{figure}

The derivation of the gas metallicity also involves an important number of systematics and sources of uncertainties, with well-known discrepancies between the use of different strong-line calibrators \citep[e.g.][]{kewley2008, lopezsanchez2012}. In this regard, to assess whether the choice of the gas-metallicity indicator affects the quantitative disagreement between the local and global laws, we alternatively derive the oxygen abundance making use of the theoretical calibrator HII-CHI-MISTRY \citep{perezmontero2014} based on photoionisation models from {\tt CLOUDY} \citep{ferland2013}. The new comparison of the slopes inferred from the local and global $Age_\star \!-Z_g$ relations as function of $M_\star$ is shown in Figure~\ref{fig5}. Consistent with the use of the O3N2 indicator, we obtain the same trend for both local and global relations of decreasing the slope when increasing the galaxy mass, with the majority of the galaxies (except the most massive ones) presenting a positive correlation between $Z_g$ and $D(4000)$. In spite of the local relations displaying larger dispersions, with HII-CHI-MISTRY we find that the quantitative agreement between the local and global relations is significantly better, with a reduction in the slope differences up to $60\%$. However, the differences are still noticeable. 

The question arises as to whether the observed differences between the global and local relations invalidate the argument that the global correlation arises from the spatial integration of the local correlation. It does not. As we show in Appendix~\ref{sec:appendix2}, the effect of random noise easily explains the difference. The error in determining stellar ages is large. Even if $D(4000)$ is well determined, it is only a proxy for stellar age with large uncertainty \citep[e.g.,][]{sanchezalmeida2012}. As soon as the errors are comparable with the range of abscissae, the slope inferred from a least squares fitting systematically underestimates the true slope. The fact that errors in spatially resolved data are larger than integrating over a galaxy naturally explains why local slopes are systematically smaller. A factor of two drop, consistent with the observations (Figure~\ref{fig4}), is easy to account for (see demonstration in Appendix~\ref{sec:appendix2}).

In this study we show the existence of both a local (spatially-resolved) and a global (galaxy-integrated) relation between $Z_g$ and $D(4000)$, that we interpret as a true relation between gas metallicity and stellar age. However, there is evidence of the influence of other physical parameters on $D(4000)$ that might affect this interpretation. One explored parameter is dust attenuation. $D(4000)$ is usually considered robust against dust-induced reddening since it is a differential measure with a relatively small wavelength separation between bands. Therefore, dust is not expected to significantly influence the relation observed between $Z_g$ and $D(4000)$, although residual effects might play a role \citep{macarthur2005}. We explore the impact of dust by investigating the shape of the local relation between dust attenuation (measured using $A_V$) and gas metallicity. We find that the majority of galaxies exhibit negative or close-to-zero slopes, in contrast to the positive slopes of the relation between $Z_g$ and $D(4000)$, suggesting therefore that the effect of dust, if existing, is minor. This is also supported by the observed trend for the dependence of the slope with the galaxy mass, with the slope of the $A_V-Z_g$ relation increasing with the galaxy mass, while in the case of the $Age_\star \!-Z_g$ relation the slope decreases with mass. Another relevant parameter to investigate is the stellar metallicity. One could think that the existence of the $Age_\star-Z_g$ relation might be partly induced by the secondary dependence of $D(4000)$ with stellar metallicity. However we believe that this is not the case, or at least, that the effect of the stellar metallicity is minor. As shown by \citet{kauffmann2003}, the dependence of $D(4000)$ on stellar metallicity is stronger for older stellar ages, being $D(4000)$ almost insensitive to it for mean ages lower than $\sim 1$ Gyr, corresponding to $D(4000) < 1.4-1.5$  ($\sim 97\%$ of the data points fall within this range). If the stellar metallicity is responsible for the observed relations, the slope of the relations should increase with the galaxy mass, which is the opposite behaviour to the observed one, in which the relations disappear for more massive systems. In conclusion, all this reasoning reinforces that the described relation between $Z_g$ and $D(4000)$ corresponds to a true relation between gas metallicity and stellar age. Since, as argued above, the strength of the 4000-\AA \,break is a measure of the relative contribution of young hot stars, it hence gives an estimate of the specific star formation rate of the galaxy \citep[e.g.][]{bruzual1983}. For this reason, the correlation between stellar age (or $D(4000)$) and gas metallicity is probably directly related to the found anti-correlation between $Z_g$ and SFR found by \citetalias{sanchezmenguiano2019}, having both of them the same physical origin.

\vspace{1cm}
In summary, this work shows for the first time the existence of a positive linear correlation between $D(4000)$ (used as an indicator of stellar age) and $Z_g$ at local scales (before and after correcting for radial trends), with a slope that decreases with increasing the galaxy stellar mass. The same trend is also reported for the global (integrated) values, confirming previous studies. We find a discrepancy in the slopes, with the local relations presenting systematically smaller values. The uncertainty in the analysed parameters can account for this difference, making the slope inferred from a least squares fitting systematically lower than the true value. Therefore, this result suggests that the observed global relation between the stellar age and the gas metallicity emerges from the spatial integration of their local correlation. The extension of this study to other global scaling relations and the search for their corresponding local counterparts can provide important information to constrain numerical simulations so to understand how galaxies form and evolve.

\acknowledgements 
We thank the anonymous referee for suggestions that allowed us to improve the paper. Thanks are also due to Tom\'as Ruiz-Lara for comments on the original manuscript. The work has been partly funded by the Spanish Ministry of Economy and Competitiveness (MINECO), projects AYA2012-31935 and AYA2016-79724-C4-2-P (ESTALLIDOS). SFS is grateful for the support of a CONACYT grant CB-285080 and FC-2016-01-1916, and funding from the PAPIIT-DGAPA-IN100519 (UNAM) project.

This project makes use of the MaNGA-Pipe3D dataproducts. We thank the IA-UNAM MaNGA team for creating it, and the ConaCyt-180125 project for supporting them.

Funding for the Sloan Digital Sky Survey IV has been provided by the Alfred P. Sloan Foundation, the U.S. Department of Energy Office of Science, and the Participating Institutions. SDSS acknowledges support and resources from the Center for High-Performance Computing at the University of Utah. The SDSS web site is \url{www.sdss.org}.

SDSS is managed by the Astrophysical Research Consortium for the Participating Institutions of the SDSS Collaboration including the Brazilian Participation Group, the Carnegie Institution for Science, Carnegie Mellon University, the Chilean Participation Group, the French Participation Group, Harvard-Smithsonian Center for Astrophysics, Instituto de Astrofísica de Canarias, The Johns Hopkins University, Kavli Institute for the Physics and Mathematics of the Universe (IPMU) / University of Tokyo, the Korean Participation Group, Lawrence Berkeley National Laboratory, Leibniz Institut f\"ur Astrophysik Potsdam (AIP), Max-Planck-Institut f\"ur Astronomie (MPIA Heidelberg), Max-Planck-Institut f\"ur Astrophysik (MPA Garching), Max-Planck-Institut f\"ur Extraterrestrische Physik (MPE), National Astronomical Observatories of China, New Mexico State University, New York University, University of Notre Dame, Observat\'orio Nacional / MCTI, The Ohio State University, Pennsylvania State University, Shanghai Astronomical Observatory, United Kingdom Participation Group, Universidad Nacional Aut\'onoma de México, University of Arizona, University of Colorado Boulder, University of Oxford, University of Portsmouth, University of Utah, University of Virginia, University of Washington, University of Wisconsin, Vanderbilt University, and Yale University.

\vspace*{0.1cm}

\software{astropy \citep{astropy2013, astropy2018},  
          Matplotlib \citep{hunter2007}
          }

\appendix
\section{Testing the robustness of the results}\label{sec:appendix3} 
In this appendix we describe several tests performed to assess if the reported results might be affected by different criteria applied along the study.  All tests show that the sample selection and analysis are robust, and the results are not contingent upon any of the examined factors.

In first place, due to the wide range of redshifts cover by the MaNGA mother sample, the physical spatial resolution of the data can vary significantly. This effect could bias the results, being the cases with no observed local correlations those corresponding to galaxies with low spatial resolution. In order to prevent this potential bias, the original sample was restricted to galaxies with $z < 0.05$ and observed with the two largest FoVs ($27.5''$ and $ 21.3''$). In spite of this, here we further investigate the effect of the spatial resolution on the results by removing from the sample those objects with $R_e$ below 3 times the seeing FWHM. The remaining 456 galaxies cover the same parameter space for the analysed local relations as the whole sample, showing no bias of the measured slopes related to spatial resolution. Although with a much smaller number statistics, restricting to galaxies with $R_e > 4 \,{\rm FWHM}$ produces the same result. 

Along the same lines, the MaNGA data show certain degree of oversampling due to the small size of the spaxels ($0.5''$) in comparison with the final spatial resolution ($\rm FWHM \simeq 2.5''$). This can induce some correlation in the data points of the local relations, which in the case of galaxies with a low number of points, might affect their determination. In this regard, we have investigated how the derived slopes are influenced by the number of points involved in their computation. We increased the threshold in the number of points requested to perform the linear fits from 30 to 100, with no significant biases in the coverage of the parameter space of the two analysed local relations. Therefore, the cases where no correlation is found do not seem to be associated with a possible oversampling of the data.

Another factor we have explored is the spatial coverage of the galaxies. As mentioned in Section~\ref{subsec:manga}, the coverage of the MaNGA galaxies is not homogeneous but some systems are observed up to $2.5-3\,R_e$, whereas others just up to $1-1.5\,R_e$. Thus, our results might overweight the centres of galaxies so that the observed behaviour might depend on the fraction of the galaxy that is covered. To address this issue, we removed from the sample those galaxies for which the FoV covers less than $2\,R_e$. Although the number of analysed objects is reduced to 224, the distribution of slopes for the local relations is similar to that of the entire sample, remaining also the reported trend with stellar mass. Therefore, the differences in the spatial coverage of the analysed sample do not seem to significantly affect the results either.

Additional tests have been performed to assess the effect of including mergers in the sample or galaxies that are too inclined to properly map the spatial variations across the discs. For the first case, we make use of the attribute {\it P\_merg}, available in the MaNGA Deep Learning Morphology Value Added catalog (MDLM-VAC) published in \citet{fischer2019} (also used for the morphological classification of the sample, see Section~\ref{subsec:sample}). This parameter measures the probability of merger signature (or projected pair). Removing from the sample 84 galaxies with {\it P\_merg} $> 0.8$, we find no significant differences in the coverage of the parameter space for the local relations. Lastly, restricting to very face-on galaxies ($i < 45 \degr$), produces no noticeable bias in the results. 

For the spaxel selection, as explained in Section~\ref{sec:analysis}, we impose a lower limit of $\rm SNR =3$ in all the emission lines involved in the derivation of the oxygen abundances, namely, $\rm H\beta$, $[\oiii] \lambda5007$, $\rm H\alpha$, and $[\nii] \lambda6584$. This restrictive limit preferentially excludes spaxels with weak emission, which may bias the local relations somehow. In order to investigate this effect, we compared the slopes of the local relations of 100 randomly selected galaxies using a SNR threshold of 1 and 3. We obtain very similar values in all cases, and a representation of one against the other shows nearly one-to-one relations (slope of 0.96 for the original relation and 1.01 for the residual relation) with very high correlation coefficients (0.94 and 0.92, respectively). This indicates that spaxels with SNR below and above 3 present similar distribution in the $D(4000)-Z_g$ parameter space, hence the values of the slopes hardly change.

In order to describe the global relation for $D(4000)$ and $Z_g$, we use the average values of all analysed star-forming regions as characteristic of the population of the entire galaxy (this is explained in Section~\ref{subsec:properties}). However, as these properties are not constant across galaxies but they present a radial decrease, their estimation might be affected by the radial extension of the galaxy covered with MaNGA data. We explored the impact of these aperture effects on the global relation (their influence on the local relation was already examined). Instead of using the average $D(4000)$ and $Z_g$ of the analysed star-forming regions, we made use of the characteristic values at $1 \, R_e$. The resulting global $Age_\star-Z_g$ relation is very similar to that shown in Figure~\ref{fig4}, with comparable slopes in all mass ranges.

Finally, for the characterisation of the local $Age_\star-Z_g$ relations we perform an ODR fitting to the data points of each individual galaxy enclosed within the 80\%- level density contour, in order to avoid outliers influencing the fits (see Section~\ref{sec:results2}). We have also tested how the use of this cut might affect the analysis. The slopes derived with and without the mentioned cut are in very good agreement with each other, for both the original and the residual relations. We obtain tight correlations very close to the one-to-one relation with high correlation coefficients ($\sim0.8$). Furthermore, the trend of the slope of the local relations with the galaxy mass show very similar behaviours to those described in Figure~\ref{fig3}, with almost imperceptible differences, concluding that the use of the cut does not change the overall results. 

\section{The $Age_\star-Z_{\lowercase{g}}$ relation from the stellar continuum fitting}\label{sec:appendix1} 

As a result of the stellar continuum fit, {\scshape Pipe3D} provides 2D luminosity-weighted (LW) stellar age maps from the individual weights of the combined SSPs (see Section~\ref{sec:analysis}). However, as explained in Section~\ref{sec:results1}, we have preferred to use $D(4000)$ as a proxy for the age of the stars, avoiding uncertainties involved in the inversion method and the complex modelling of the stellar populations applied to spatially resolved IFS data \citep[e.g.][]{cidfernandes2014, sanchez2016a, bittner2019}. Nevertheless, for the sake of comprehensiveness, this appendix shows the comparison of the corresponding local and global $Age_\star-Z_g$ relations using the 2D LW stellar age maps provided by {\scshape Pipe3D}.

\begin{figure}
\begin{center}
\resizebox{\hsize}{!}{\includegraphics{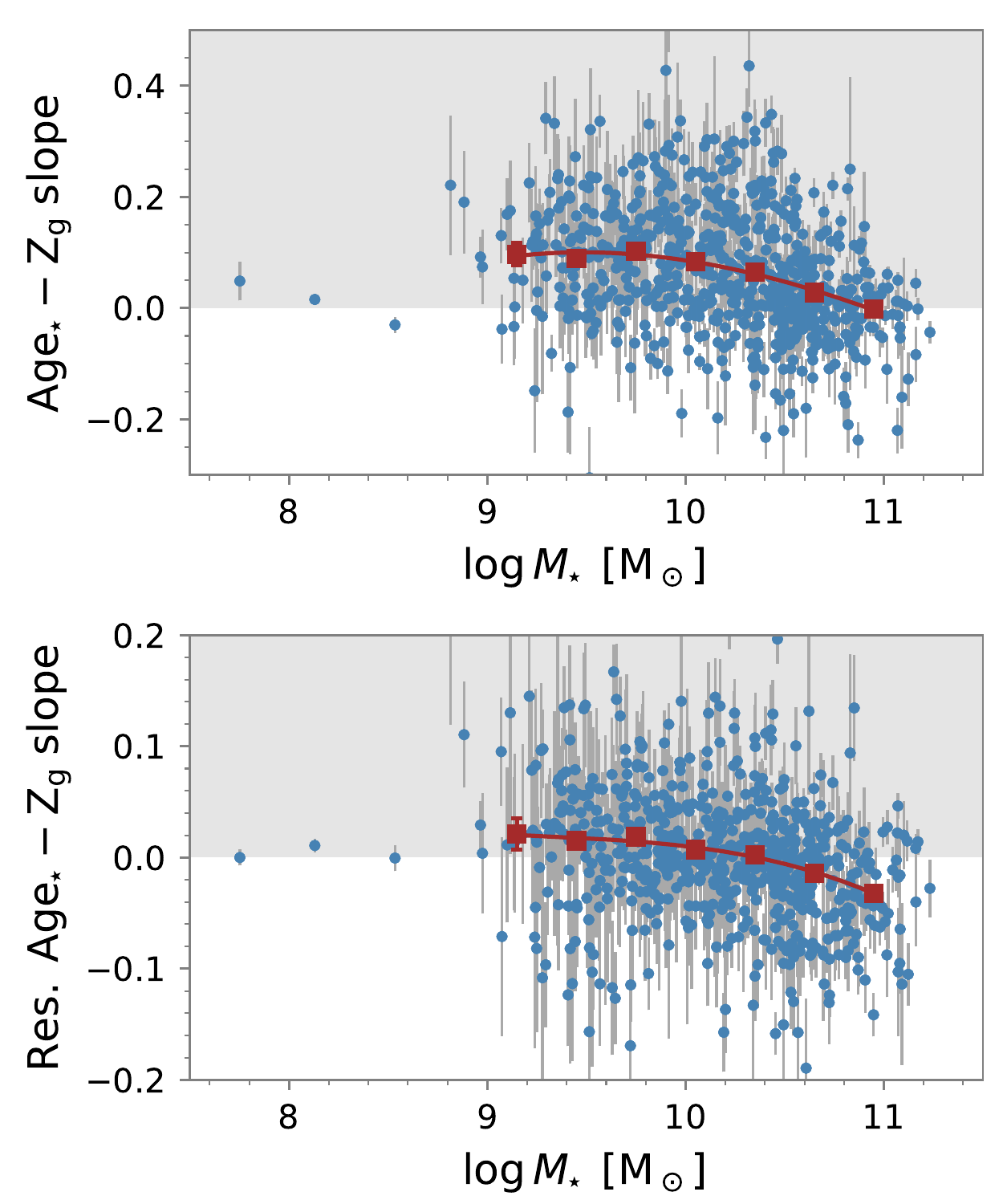}}
\caption{Slope of the $Age_\star-Z_g$ local relations as function of $M_\star$ (original, {\it top}; residual, {\it bottom}). $Age_\star$ is derived from the stellar continuum fit. See caption of Figure~\ref{fig3} for more details.}
\label{fig1_ap1}
\end{center}
\end{figure}

The local relations between $Age_\star$ and $Z_g$ are derived for all the galaxies in the sample. We find a positive (negative) slope for 58\% (15\%) of the galaxies. The remaining 27\% are compatible with absence of correlation (i.e., zero slope). When considering the variations of the residuals $\Delta Age_\star$ and $\Delta Z_g$, just 27\% (28\%) of the galaxies show positive (negative) correlation, with the 45\% of the sample showing lack of correlation within the observational errors. The high presence of no correlations is probably due, as mentioned in Appendix~\ref{sec:appendix2}, to the typical uncertainty associated with age determinations from spectral fitting, since the percentage of slopes compatible with zero is significantly smaller (22\%) when using $D(4000)$. 

Figure~\ref{fig1_ap1} shows the slope of the original (top) and residual (bottom) local relations as a function of the galaxy mass (see Section~\ref{sec:results2} for details on the layout of the figure). For the original relation (top panel), we see that lower mass galaxies present a larger slope than more massive galaxies, where the slope tends to zero and the correlation disappears. In the case of the residual local relation (bottom panel), the slopes are very small for most galaxies, although again the least massive galaxies tend to display the largest slopes. Similarly to $D(4000)$, the relation reverses at $\rm \log (M_\star/M_\odot) \sim10-10.5$, with more massive galaxies showing an anti-correlation between the variations of the residuals $\Delta Age_\star$ and $\Delta Z_g$.

\begin{figure}
\begin{center}
\resizebox{\hsize}{!}{\includegraphics{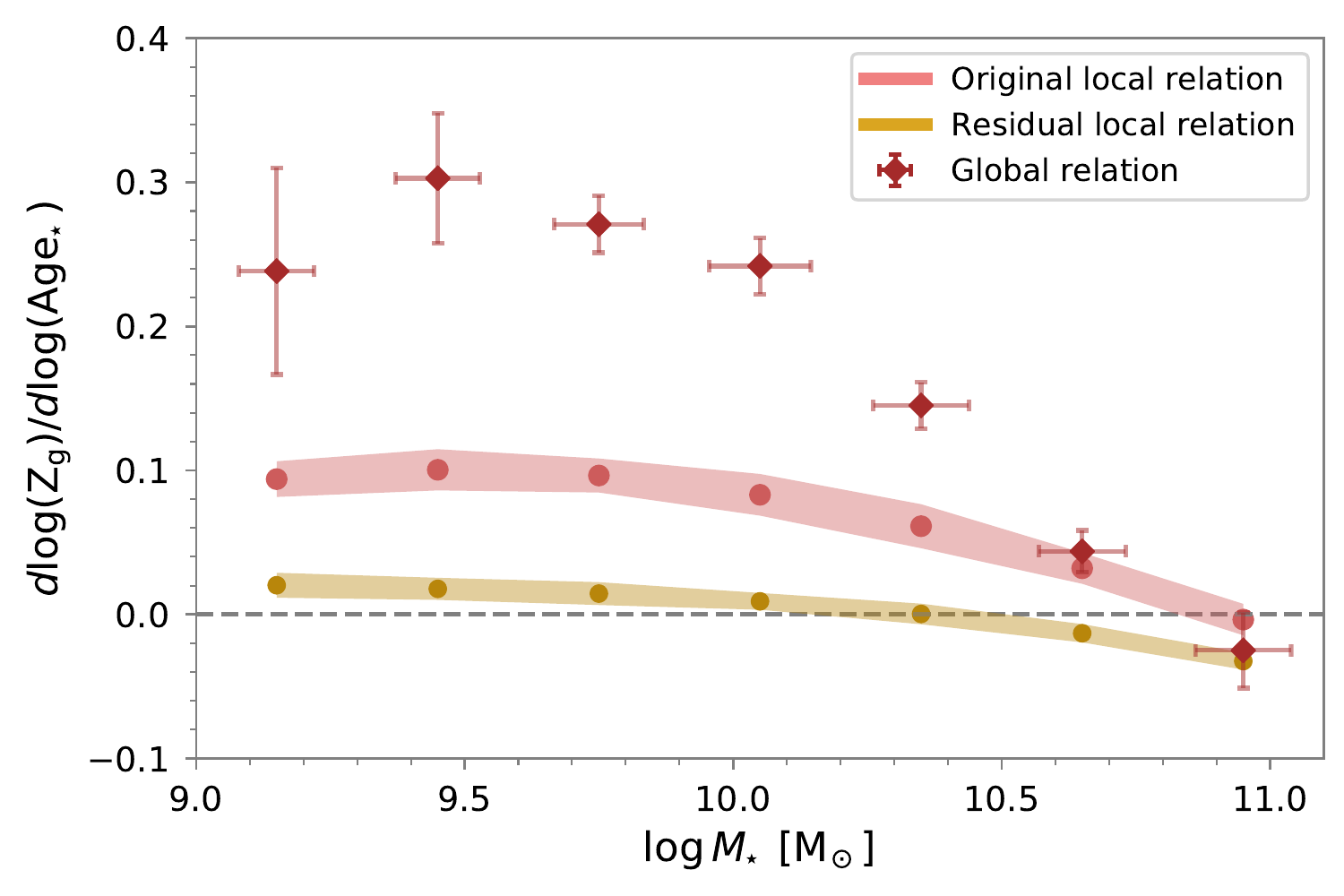}}
\caption{Comparison of the slope inferred from local and global $Age_\star \!-Z_g$ relations as a function of $M_\star$. $Age_\star$ is derived from the stellar continuum fit. See caption of Figure~\ref{fig4} for more details.}
\label{fig2_ap1}
\end{center}
\end{figure}

Finally, Figure~\ref{fig2_ap1} represents the comparison of the slopes derived from the global and local relations at each M$_\star$ (see Section~\ref{sec:localglobal} for details on the layout of the figure). For the global law, we use the mean LW stellar age value measured for all the analysed bins associated with star formation as characteristic of the population of the entire galaxy (see Section~\ref{sec:results2} for details on the binning scheme).  We can see that, overall, the three relations show the same tendency of decreasing slope when increasing galaxy mass. In spite of the similarity in the tendency, the relations do not agree quantitatively; the global relation reach much more positive slopes than the local ones in the low mass regime, with the residual local relation presenting the values closest to zero. As we argue in Appendix~\ref{sec:appendix2}, the presence of noise can explain these differences between local and global correlations, as well as the difference with the correlations found when using $D(4000)$ as age indicator (Section~\ref{sec:localglobal}).

\section{Random noise explains the difference of slopes between local and global relations}\label{sec:appendix2}

Suppose that two variables, $x$ and $y$, are related linearly so that
\begin{equation}
  y = m x + k.
  \label{eq:linear}
\end{equation}
We observe them in $N$ different points with some error, i.e.,
\begin{eqnarray}
  x_{io}=&x_i+\Delta x_i,\nonumber \\
  y_{io}=&y_i+\Delta y_i,
           \label{eq:def}
\end{eqnarray}
where $x_{io}$ and $y_{io}$ are the $i$-th observation, $x_i$ and $y_i$ follow the law in Eq.~(\ref{eq:linear}), and $\Delta x_i$ and $\Delta y_i$ are the error in the two observed variables.  $\Delta x_i$ and $\Delta y_i$ are assumed to be random independent variables of mean zero and variance $\sigma_x^2$ and $\sigma_y^2$, respectively. We try to infer the slope $m$ through the customary least-squares fit, i.e.,
\begin{equation}
m_{ls} \simeq  \frac{N\,\sum_i x_{io}y_{io}- \sum_ix_{io}\,\sum_iy_{io}}{N\sum_ix_{io}^2- (\sum_ix_{io})^2}.
\label{eq:slopels}
\end{equation}
Using Eq.~(\ref{eq:def}) and neglecting all the sums expected to average to zero (i.e., those having $\Delta x_i$, $\Delta y_i$, $y_i \Delta x_i$, $x_i\Delta y_i$, $x_i\Delta x_i$, and $\Delta x_i\,\Delta y_i$), one ends up with an expression for the least squares slope $m_{ls}$ in terms of the true slope $m$,
\begin{equation}
m_{ls} =\frac{m}{1+(\sigma_x/{\rm RMS}_x)^2},
\end{equation}
with ${\rm RMS}_x$ defined as
\begin{equation}
  {\rm RMS}_x^2 = \frac{1}{N}\sum_i\Big(x_i-\Sigma_j x_j/N\Big)^2.
  \end{equation}
${\rm RMS}_x$ is the typical range of abscissae used in the fit. Thus, as soon as the range of abscissae is comparable to their error (${\rm RMS}_x\sim \sigma_x$), the least squares based slope is significantly smaller than the true slope ($m_{ls}\sim m/2$). As an example, the data points in Figure~\ref{fig2} have $\sigma_x/{\rm RMS}_x$ of the order of 0.89 and 0.86 (central and right panels, respectively). The mean value of the ratio considering 100 randomly selected galaxies is 0.8. $\sigma_x$ was estimated from the observed points as the standard deviation of the observed $D(4000)$ (or $\Delta D(4000)$) minus the value predicted by the linear fit. ${\rm RMX}_x$ was estimated as the standard deviation of the $D(4000)$ (or $\Delta D(4000)$) predicted by the linear fit. The fact that the errors are larger in the spatially resolved data than when integrating over a galaxy naturally explains why local slopes are systematically smaller.    

\bibliographystyle{aasjournal}
\bibliography{bibliography}

\end{document}